\documentclass[twocolumn,showpacs,amsmath,amssymb]{revtex4}
\usepackage{graphics}
\usepackage{epsfig}

\begin{document}

\title{Measurement of the Negative Muon Anomalous Magnetic Moment to 0.7~ppm}

\author{
G.W.~Bennett$^{2}$,
B.~Bousquet$^{9}$,
H.N.~Brown$^2$,
G.~Bunce$^2$,
R.M.~Carey$^1$,
P.~Cushman$^{9}$,
G.T.~Danby$^2$,
P.T.~Debevec$^7$,
M.~Deile$^{11}$,
H.~Deng$^{11}$,
S.K.~Dhawan$^{11}$,
V.P.~Druzhinin$^3$,
L.~Duong$^{9}$,
F.J.M.~Farley$^{11}$,
G.V.~Fedotovich$^3$,
F.E.~Gray$^7$,
D.~Grigoriev$^3$,
M.~Grosse-Perdekamp$^{11}$,
A.~Grossmann$^6$,
M.F.~Hare$^1$,
D.W.~Hertzog$^7$,
X.~Huang$^1$,
V.W.~Hughes$^{11}$\footnote[2]{Deceased.},
M.~Iwasaki$^{10}$,
K.~Jungmann$^5$,
D.~Kawall$^{11}$,
B.I.~Khazin$^3$,
F.~Krienen$^1$,
I.~Kronkvist$^{9}$,
A.~Lam$^1$,
R.~Larsen$^2$,
Y.Y.~Lee$^2$,
I.~Logashenko$^{1,3}$,
R.~McNabb$^{9}$,
W.~Meng$^2$,
J.P.~Miller$^1$,
W.M.~Morse$^2$,  
D.~Nikas$^2$, 
C.J.G.~Onderwater$^{7}$,
Y.~Orlov$^4$,
C.S.~\"{O}zben$^{2,7}$,
J.M.~Paley$^1$,   
Q.~Peng$^1$,   
C.C.~Polly$^7$,   
J.~Pretz$^{11}$,   
R.~Prigl$^{2}$,
G.~zu~Putlitz$^6$,
T.~Qian$^{9}$,  
S.I.~Redin$^{3,11}$,
O.~Rind$^1$,
B.L.~Roberts$^1$,
N.~Ryskulov$^3$,
Y.K.~Semertzidis$^2$, 
P.~Shagin$^9$,
Yu.M.~Shatunov$^3$,
E.P.~Sichtermann$^{11}$,
E.~Solodov$^3$,
M.~Sossong$^7$, 
L.R.~Sulak$^{1}$,
A.~Trofimov$^1$,
P.~von~Walter$^6$,
and
A.~Yamamoto$^8$.
\\
(Muon $(g-2)$ Collaboration)
}

\affiliation{
\mbox{$\,^1$Department of Physics, Boston University, Boston, Massachusetts 02215}\\
\mbox{$\,^2$Brookhaven National Laboratory, Upton, New York 11973}\\
\mbox{$\,^3$Budker Institute of Nuclear Physics, Novosibirsk, Russia}\\
\mbox{$\,^4$Newman Laboratory, Cornell University, Ithaca, New York 14853}\\
\mbox{$\,^5$ Kernfysisch Versneller Instituut, Rijksuniversiteit Groningen, NL 9747\,AA Groningen, The Netherlands}\\
\mbox{$\,^6$ Physikalisches Institut der Universit\"at Heidelberg, 69120 Heidelberg, Germany}\\
\mbox{$\,^7$ Department of Physics, University of Illinois at Urbana-Champaign, Illinois 61801}\\
\mbox{$\,^8$ KEK, High Energy Accelerator Research Organization, Tsukuba, Ibaraki 305-0801, Japan}\\
\mbox{$\,^{9}$Department of Physics, University of Minnesota,Minneapolis, Minnesota 55455}\\
\mbox{$\,^{10}$ Tokyo Institute of Technology, Tokyo, Japan}\\
\mbox{$\,^{11}$ Department of Physics, Yale University, New Haven, Connecticut 06520}
}

\begin{abstract}
The anomalous magnetic moment of the negative muon has been measured
to a precision of 0.7 parts per million (ppm) at the Brookhaven
Alternating Gradient Synchrotron.  This result is based on data 
 collected in 2001, and  is over an order of
magnitude more precise than the previous measurement of the negative muon.  
The result
$a_{\mu^-} = 11\,659\,214 (8)(3) \times 10^{-10}$ (0.7\,ppm), where
the first uncertainty is statistical and the second is systematic, is 
consistent with previous measurements of the anomaly for the
positive and negative muon.  The average for
the muon anomaly is $a_\mu(\text{exp}) = 11\,659\,208 (6) \times
10^{-10}$ (0.5\,ppm).
\end{abstract}
\pacs{13.40.Em, 12.15.Lk, 14.60.Ef}

\maketitle

The anomalous magnetic moments of the muon
and the electron have played an important role in the
development of the standard model.  Compared to the electron, the muon
anomaly has a relative sensitivity to heavier mass 
scales which typically is proportional
to $(m_{\mu} /m_e)^2$. At the present level of accuracy 
the muon anomaly gives an
experimental sensitivity to virtual $W$ and $Z$ gauge bosons as well
as a potential sensitivity to other, as yet unobserved, particles in
the few hundred GeV/$c^2$ mass range~\cite{kh}. 

  We report our result for the  negative muon anomalous
  magnetic moment $a_{\mu^-}=(g-2)/2$ from data collected in early 2001.  
  The measurement is based on muon spin precession in
  a magnetic storage  ring with electrostatic focusing.
 The same experimental technique was used 
  as in our most recent  measurements of  $a_{\mu^+}$~\cite{muplus,g2_2000}, and  a similar
  precision of 0.7 ppm was achieved.  Detailed descriptions of the apparatus may be
  found elsewhere~\cite{nimpapers,fei,nimpapers2,kicker,quads}.

For polarized muons moving in a uniform magnetic field $\vec{B}$ perpendicular
to the muon spin  and to the plane of the orbit and in an
electric quadrupole field $\vec{E}$, which is used for vertical focusing~\cite{quads},
the angular frequency difference, $\omega_a$ between the spin precession frequency
 and the cyclotron frequency, is given by
\begin{equation}
\vec{\omega}_a= {e \over m \, c}  \left[ a_\mu \vec{B} - \left( a_\mu - {1 \over \gamma^2 - 1} \right) \vec{\beta} \times \vec{E} \right].
  \label{eq:amu}
\end{equation}
The dependence of $\omega_a$ on the electric field is eliminated by storing 
muons with the ``magic'' $\gamma=29.3$~\cite{cern}, which corresponds to a 
muon 
momentum $p=3.09$~GeV/$c$.  Hence measurement of $\omega_a$ and of $B$, 
in terms of the free proton NMR frequency $\omega_p$ and the ratio of 
muon to proton magnetic moments $\lambda$, determines 
$a_\mu$.  At the magic $\gamma$, the muon lifetime is approximately
 $ 64.4 \, {\rm \mu s}$ and the $(g-2)$ precession period is 
$4.37\, {\rm \mu s}$.  With a field
of 1.45~T in our storage ring~\cite{nimpapers}, the central orbit 
radius is 7.11~m.
%
%
%
%
%

The difference frequency
$\omega_a$ was determined by counting the number $N(t)$
of decay electrons above an 
energy threshold.  The time
spectrum of decay electrons is then given by
\begin{equation}
  N(t) = N_0 e^{-t/(\gamma\tau)} \left[ 1 + A \sin(\omega_a t + \phi_a) \right].
  \label{eq:spectrum}
\end{equation}
The normalization $N_0$, asymmetry $A$, and phase
$\phi_a$ vary with the chosen energy threshold.  
%

\begin{figure}
  \includegraphics[width=0.45\textwidth,angle=0]{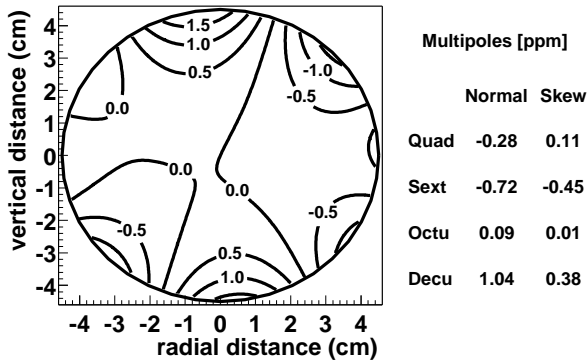}
  \caption{A two-dimensional multipole expansion of the 2001 field averaged
           over azimuth from one out of 20 trolley measurements.  Half
           ppm contours with respect to a central azimuthal average
           field \mbox{$B_0 = 1.451\,269\,\text{T}$} are shown.  The
           multipole amplitudes relative to $B_0$ are given at the
           beam aperture, which had a radius of 4.5\,cm and is
           indicated by the circle.\label{fig:field}}
\end{figure}

The measurement of the magnetic field frequency $\omega_p$ is based on proton NMR in water.
A trolley with 17 NMR probes was moved typically every three days throughout the entire muon storage region.
About 150 fixed NMR probes distributed around the ring in
the top and bottom walls of the vacuum chamber were used to
interpolate the field between trolley measurements.  The system was
calibrated with respect to a standard probe with a spherical
\mbox{$\text{H}_2\text{O}$} sample~\cite{fei}.
The homogeneity of the  field in 2001 (Figure~\ref{fig:field}) was similar to 
that  achieved for the opposite polarity field in 2000~\cite{g2_2000}.

The field $\langle B \rangle$ weighted with the analyzed event sample
was obtained from two largely independent analyses, whose results were
found to agree to within 0.05\,ppm.  Its final value is expressed in
terms of the free proton resonance frequency and is given by
\mbox{$\omega_p/(2\pi) = 61\,791\,400(11)\,\text{Hz}$ (0.2\,ppm).}
Table~\ref{table:field} lists the uncertainties.  The improved 2001 uncertainties
resulted from refinements in the calibration measurements, and from an
upgraded system to determine the azimuthal trolley position in the
storage ring.
\newsavebox{\tempbox}
\sbox{\tempbox}{
  \begin{minipage}{0.45\textwidth}
    \begin{small}
      $^\dagger$ higher multipoles, trolley temperature and voltage response,
      eddy currents from the kickers, and time-varying stray fields.
    \end{small}
  \end{minipage}
}
\begin{table}
\caption {Systematic uncertainties for the $\omega_p$ analysis.
          \label{table:field}}
\begin{tabular}{l|c}
\hline\hline
Source of uncertainty & Size [ppm] \\
\hline
Absolute calibration of standard probe\hspace{3em} & 0.05\\
Calibration of trolley probe & 0.09\\
Trolley measurements of $B_0$ & 0.05\\
Interpolation with fixed probes & 0.07\\
Uncertainty from muon distribution & 0.03\\
Others$^\dagger$ & 0.10\\
\hline
Total systematic error on $\omega_p$ & 0.17 \\
\hline\hline
\end{tabular}
\usebox{\tempbox}
\end{table}

\begin{figure}
  \includegraphics[width=0.4\textwidth,angle=0]{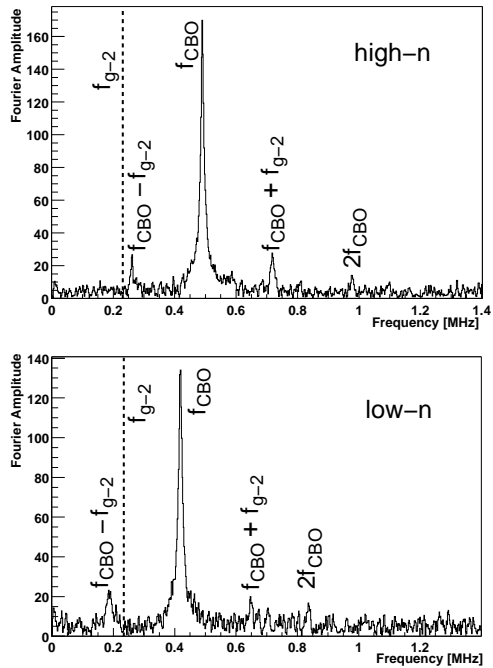}
  \caption{The Fourier spectrum of the residuals of a fit to the 
           five free parameters given in Eq.~\ref{eq:spectrum} for
           the high (top) and low (bottom) $n$-value data.  The corresponding
           CBO frequencies, located at
           491~kHz (top), and 419~kHz (bottom) 
           as well as their $(g-2)$ sidebands are clearly visible. 
           Dashed lines indicate the $(g-2)$ frequency. 
           \label{fig:FT_high_low}}
\end{figure}

The 2001 $\omega_a$ data taking was similar to that in 2000.
However, the hardware energy threshold of
the detectors was kept lower and equal for all counters at  0.9\,GeV compared 
to 1.0-1.4~GeV in 2000.  This was made
possible by reducing the intensity of the injected beam, which in turn
reduced the light flash in the detectors~\cite{muplus,g2_2000}. 
  These factors allowed all the detectors to be turned on and be stable
by 32\,$\mu$s after beam injection, as opposed to
50\,$\mu$s in 2000.  As a result of the reduced rates, the fraction of overlapping
signals (pileup) after 32\,$\mu$s in 2001 was comparable to the pileup fraction
after  50\,$\mu$s in 2000.
 In 2000 
the field focusing index $n$, which is proportional to the electric field gradient, was $n=0.137$, corresponding to a 
horizontal coherent betatron oscillation frequency (CBO) of 466~kHz~\cite{g2_2000}.
  This frequency was close to 
twice the $(g-2)$ frequency of  229~kHz, which resulted in a sizable 
uncertainty in the fitted $\omega_a$ value~\cite{g2_2000}.  In 2001 we 
used two 
different $n$-values, $n = 0.122$ and $n= 0.142$, which resulted in CBO 
frequencies, 419~kHz and 491~kHz
 that are further  from twice the 
$(g-2)$ frequency (see  Figure~\ref{fig:FT_high_low}). 
 Consequently, the uncertainty caused by CBO is 
smaller. Furthermore it also reduced the correlation between the CBO and 
detector gain effects
in the fits to the time spectrum.

Two independent
implementations of the algorithm to 
reconstruct the electron times and energies from the calorimeter 
signals were used.
The frequency $\omega_a$ was determined by fitting the time 
distribution of decay electrons. 
Five independent analyses were performed in order to probe the 
systematic uncertainties and, of course, to protect against mistakes. 
All five results agreed within the expected statistical 
deviations due to different data selection and weightings. These analyses
are described below.

Two of the analyses used slightly different
parametrizations~\cite{muplus,g2_2000} that included 
CBO modulations and fitted the combined electron spectrum in the energy 
range  1.8-3.4 GeV.  
In the third analysis, the counts were weighted with the experimentally
determined energy-dependent modulation asymmetry, which optimized the
statistical power of the data.  This method permitted the analyzed
energy range to be extended.  We used an energy range of 1.5 to 3.4~GeV,
which together with the asymmetry weighting
resulted in a 10\% improvement of the statistical uncertainty.   
As in the first two
analyses, the resulting spectrum of {\em weighted} counts was fitted
to a function that parametrized all known and statistically
significant perturbations.

The remaining  analyses fit the ratio~\cite{muplus,g2_2000}
 formed by randomly assigning
the data to four statistically independent subsets $n_1$ to $n_4$.  
The subsets were rejoined in $u(t)=n_1(t) + n_2(t)$ and
$v(t)=n_3(t-\tau_a / 2) + n_4(t + \tau_a /2)$,
where $\tau_a$ is an estimate of the $(g-2)$ period, and then
combined to form the time spectrum 
$ r(t) = [{u(t)-v(t)}]/[{u(t)+v(t)}]$.  The $(g-2)$ rate modulation of
$v$ is 180$^\circ$ degrees out of phase compared to that of 
$u$, and to sufficient precision $r(t)$ can be described by $A\sin(\omega_a
t + \phi_a)$.  The ratio $r(t)$ is largely insensitive to changes of observed
counts on time scales larger than $\tau_a = 2\pi/\omega_a \sim
4\,\mu\mathrm{s}$.

In one of the  ratio analyses, the sensitivity to CBO was reduced
by combining the data from both $n$-values and all detectors prior to
fitting.
The data were fitted from 32\,$\mu$s after injection when all detectors were on.
In the second ratio 
analysis the data were fitted separately for each calorimeter and $n-$value.
The fits began between 24\,$\mu$s and 32\,$\mu$s, and required the 
parametrization of the CBO effects in the fit function.

%
%
%
%
%

Changes in the radial and vertical muon distributions with time were 
quantified, and were found to have
negligible effect on $\omega_a$.
A small reduction in the pulsed electrostatic quadrupole voltages~\cite{quads}  
during the measurement period could change the vertical muon 
distribution.
 Analysis of the data from scintillator counter
hodoscopes placed in front of the calorimeters
combined with a beam tracking calculation and a
 GEANT based simulation set a systematic error limit of
 0.03~ppm.  The muon radial distribution is determined by the
magnetic field and the momentum distribution~\cite{muplus,yuri}.  The magnetic field
 does not change with time after injection, except due to the
field from eddy currents induced by the fast kicker~\cite{kicker}. 
This was measured, and found to have a negligible effect on the 
muon radial
distribution. Muons of lower momenta decay earlier in the laboratory frame than 
muons of higher momenta.  The momentum distribution of the stored beam 
thus changes during the $~600\, \rm \mu s$ measurement period.  The
effect on $\omega_a$ due to this change was studied in simulation, and
was found to be  0.03~ppm.  

%
\begin{figure}
  \includegraphics[width=0.45\textwidth,angle=0]{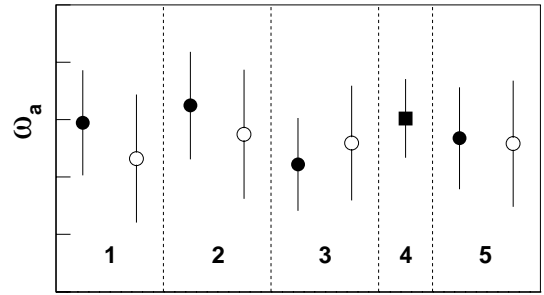}
  \caption{Comparison of the $\omega_a$ values from the five analyses 
for the low-$n$ (filled)
and high-$n$ (open) data sets.  Analysis 4 used only the combined low 
and high-$n$
data (square). The
divisions on the vertical axis are separated by 1 ppm, and
the indicated uncertainties are statistical.  The systematic 
uncertainties are considerably smaller.
}
  \label{fig:all_results}
\end{figure}

The results for $\omega_a$ for the two $n$-values are consistent, see
Figure~\ref{fig:all_results}, and were combined for each of the analyses.
 The values for $\omega_a$ from the five analyses are in 
agreement to within variations expected from the differences in the 
analyzed event samples and the treatment of the data. The analysis 
techniques are expected to have somewhat different sensitivities to 
different systematic effects.
Detailed comparisons of the results, using all analyzed data as well as 
only the data in overlap, showed no evidence for unaccounted systematic 
differences.  The five resulting values for $\omega_a$ were combined in a 
simple arithmetic mean to obtain a single value for $\omega_a$.

The resulting frequency value is 
$\omega_a/(2\pi) = 229\,073.59(15)(5)\,\mathrm{Hz}$ (0.7\,ppm), 
which includes a correction of $+0.77(6)$ ppm for
contributions to Eq.~\ref{eq:amu} caused by vertical oscillations (0.30\,ppm)
and for the effect of the horizontal electric fields on
muons with \mbox{$\gamma \neq 29.3$} (0.47\,ppm).
The stated uncertainties account for strong correlations among the 
individual results, both statistical and systematic.
Table~\ref{table:omegaa} lists the  systematic uncertainties in the 
combined result with these correlations taken into account.
\sbox{\tempbox}{
  \begin{minipage}{0.45\textwidth}
    \begin{small}
    \noindent$^\dagger$ AGS background, timing shifts, E field and 
vertical oscillations, beam debunching/randomization, 
binning and fitting procedure.
    \end{small}
  \end{minipage}
}
\begin{table}
\caption{Systematic uncertainties for the combined $\omega_a$ analysis.}
\begin{tabular}{l|c}
\hline\hline
Source of errors & Size [ppm] \\
\hline
Coherent betatron oscillations\hspace{6em}  &  0.07\\
Pileup & 0.08\\
Gain changes & 0.12\\
Lost muons & 0.09\\
Others$^\dagger$ & 0.11\\
\hline
Total systematic error on $\omega_a$ & 0.21\\
\hline\hline
\end{tabular}
\usebox{\tempbox}
\label{table:omegaa}
\end{table}

\begin{figure}[h!]
  \includegraphics[width=0.45\textwidth,angle=0]{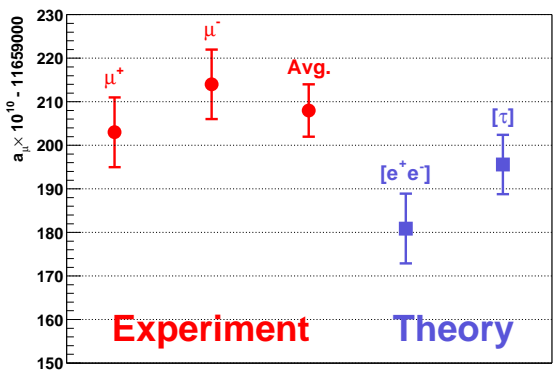}
  \caption{Measurements of $a_\mu$ by E821 with the SM predictions (see text
for discussion). Uncertainties indicated on the measurements are total
uncertainties.
}
  \label{fig:results}
\end{figure}

After the $\omega_p$ and $\omega_a$ analyses 
were finalized separately and independently, $a_\mu$ was evaluated.
The result is
\begin{equation}
  a_{\mu^-} = \frac{R}{\lambda - R} = 11\,659\,214(8)(3)\ \times\ 10^{-10}~~\mbox{(0.7\,ppm)},
  \label{eq:result}
\end{equation}
where $R_{\mu^-} \equiv \omega_a/\omega_p = 0.003 \, 707 \, 208 \, 3 (2\,6) $ and 
$\lambda = \mu_\mu/\mu_p = 3.183\,345\,39(10)$~\cite{liu}.
This new result is in good agreement with the average of
$R_{\mu^+}  = 0.003\, 707\, 204\, 8 (2\,5) $~\cite{muplus}
 as predicted by the
CPT theorem.  The difference $\Delta R = R_{\mu^-} - R_{\mu^+} = (3.5 \pm 3.4)\times 10^{-9}$.
The new  average is $R_{\mu}  = 0.003\,707\,206\,3 (2\,0) $ and
\begin{equation}
  a_\mu(\mathrm{exp}) = 11\,659\,208(6) \times 10^{-10}~~\mbox{(0.5\,ppm)},
\end{equation}
in which the total uncertainty consists of $5 \times 10^{-10}$ (0.4\,ppm) 
statistical uncertainty
and $4 \times 10^{-10}$ (0.3\,ppm) systematic uncertainty.
The correlation of systematic uncertainties between the data sets has been taken into account.
The combined result for the positive muon  \cite{g2_2000}, 
$a_{\mu^+}(\mathrm{exp}) = 11\, 659\, 203(8) \times 10^{-10}$ (0.7 ppm) 
has a statistical uncertainty of $6  \times 10^{-10}$
(0.6 ppm) and a systematic uncertainty of $5 \times 10^{-10}$ (0.4 ppm).  
It is shown in Figure~\ref{fig:results} together with the 
new result for the negative muon and their average.

The standard model prediction for $a_{\mu}$ consists of QED, hadronic and
weak  contributions.  The uncertainty on the
standard model value is dominated by the 
uncertainty on the lowest-order hadronic vacuum polarization.
This contribution 
can be determined directly from the annihilation of $e^+e^-$ to
hadrons through a dispersion integral~\cite{disp}.  
The indirect determination
using data from hadronic $\tau$ decays, the conserved vector current
hypothesis, plus the appropriate isospin corrections, 
could in principle improve 
the precision of $a_{\mu}({\rm had})$.  However,
discrepancies between the $\tau$ and the $e^+e^-$ results
 exist~\cite{dehz03,jeg}.
The two data sets do not give consistent results for the pion form factor.
 Using
$e^+e^-$ annihilation data the corresponding theoretical value is
$a_\mu(\mathrm{SM}) = 11\,659\,181(8)\,\times\,10^{-10} \mathrm{\ (0.7\,ppm)}$.
The value deduced from $\tau$ decay is larger by
$15 \times 10^{-10}$ and has a stated 
uncertainty of $7\,\times\,10^{-10} \mathrm{\ (0.6\,ppm)}$.
The difference between the experimental determination of $a_\mu$ and the standard model theory
using  
the $e^+e^-$ or $\tau$  data for the calculation of the hadronic vacuum polarization is
2.7~$\sigma$ and  1.4~$\sigma$, respectively.

This is the final analysis of the anomalous magnetic moment 
from experiment E821 at the Brookhaven 
Alternating Gradient Synchrotron.  We aim to
 substantially improve our result in a 
new measurement and look forward to continued
efforts to improve the theoretical evaluation.

We thank T.~Kirk, D.I.~Lowenstein, P.~Pile, and the staff of the 
BNL AGS for the strong support they have given this experiment.
This work was supported in part by the U.S. Department of Energy,
 the U.S. National Science Foundation, the U.S. National
 Computational Science Alliance, the German Bundesminister
 f\"{u}r Bildung und Forschung, the Russian Ministry of Science,
 and the U.S.-Japan Agreement in High Energy Physics.

\end{document}